\documentclass[sn-mathphys,Numbered]{sn-jnl}

\usepackage{graphicx}
\usepackage{multirow}
\usepackage{amsmath,amssymb,amsfonts}
\usepackage{amsthm}%
\usepackage{mathrsfs}%
\usepackage[title]{appendix}%
\usepackage{textcomp}%
\usepackage{manyfoot}%
\usepackage{booktabs}%
\usepackage{algorithm}%
\usepackage{algorithmicx}%
\usepackage{algpseudocode}%
\usepackage{listings}%
\usepackage{adjustbox}

\usepackage{subfigure}
\usepackage[normalem]{ulem}
\usepackage{gensymb} 
\usepackage{upgreek}
\usepackage{float}
\usepackage{psfrag}
\usepackage{picture,xcolor}

\pdfoutput=1

\usepackage{tikz}

\usepackage{listings}

\expandafter\ifx\csname package@font\endcsname\relax\else
 \expandafter\expandafter
 \expandafter\usepackage
 \expandafter\expandafter
 \expandafter{\csname package@font\endcsname}%
\fi
\usepackage{tikz}

\usepackage{listings}

\expandafter\ifx\csname package@font\endcsname\relax\else
 \expandafter\expandafter
 \expandafter\usepackage
 \expandafter\expandafter
 \expandafter{\csname package@font\endcsname}%
\fi
\begin{document}
\title{Modification of Jet Velocities in an Explosively Loaded Copper Target with a Conical Defect}

\author[1]{\fnm{Michael P.} \sur{Hennessey}}\email{hennessey2@llnl.gov}

\author[2]{\fnm{Finnegan} \sur{Wilson}}\email{fwilson@mines.edu}

\author[2]{\fnm{Grace I.} \sur{Rabinowitz}}\email{grabinowitz@mines.edu}

\author[2]{\fnm{Max J.} \sur{Sevcik}}\email{msevcik@mines.edu}

\author[2]{\fnm{Kadyn J.} \sur{Tucker}}\email{kjtucker@mines.edu}

\author[1]{\fnm{Dylan J.} \sur{Kline}}\email{kline11@llnl.gov}

\author[1]{\fnm{David K.} \sur{Amondson}}\email{amondson1@llnl.gov}

\author[1]{\fnm{H. Keo} \sur{Springer}}\email{springer12@llnl.gov}

\author[1]{\fnm{Kyle T.} \sur{Sullivan}}\email{sullivan34@llnl.gov}

\author*[2,3]{\fnm{Veronica} \sur{Eliasson}}\email{eliasson@mines.edu}

\author[1]{\fnm{Jonathan L.} \sur{Belof}}\email{belof1@llnl.gov}

\affil[1]{\orgdiv{Physical and Life Sciences Directorate}, \orgname{Lawrence Livermore National Laboratory}, \orgaddress{\street{7000 East Avenue}, \city{Livermore}, \postcode{94550}, \state{CA}, \country{USA}}}

\affil[2]{\orgdiv{Mechanical Engineering Department}, \orgname{Colorado School of Mines}, \orgaddress{\street{1500 Illinois Street}, \city{Golden}, \postcode{80401}, \state{Golden}, \country{United States}}}

\affil[3]{\orgdiv{Mining Engineering Department}, \orgname{Colorado School of Mines}, \orgaddress{\street{1500 Illinois Street}, \city{Golden}, \postcode{80401}, \state{Golden}, \country{United States}}}


\abstract{In this work, the design and execution of an experiment with the goal of demonstrating control over the evolution of a copper jet is described. Simulations show that when using simple multi-material buffers placed between a copper target with a conical defect and a cylinder of high-explosive, a variety of jetting behaviors occur based on material placement, including both jet velocity augmentation and mitigation. A parameter sweep was performed to determine optimal buffer designs in two configurations. Experiments using the optimal buffer designs verified the effectiveness of the buffer and validated the modeling.}



\maketitle


\begin{section}{Introduction}\label{sec:introduction}

    The formation of jets in explosively loaded metals with imperfect surfaces is a well-known behavior which occurs in a range of applications including shaped charges, explosive welding, and others. These jets are believed to arise from the Richtmyer-Meshkov instability (RMI) as a limiting case.\\ 

    \noindent In general terms, RMI is a phenomenon in which a perturbed interface between two fluids of different properties deforms following an impulsive impact such as shock loading~\cite{Richtmyer1960, Meshkov1969, Brouillette2002, Zhou2017_1, Zhou2017_2, Zhou2021}. In general, shocked material flows like a fluid (hydrodynamically). The perturbation at the interface will evolve and, depending upon the materials that define the interface and the impact magnitude, may do so unbounded. The mechanism by which these initial perturbations grow is the mismatch of the shock's pressure gradient and the local density gradient across the interface, resulting in baroclinic vorticity generation. In the case of the impact traveling through a denser material into a less dense material, the desner fluid flows with the generated vortices resulting in the inversion and subsequent growth of the interface perturbations. A comprehensive review of the physical phenomena involved in the generation and study of RMI may be found in Brouillette et al.~\cite{Brouillette2002} and in the studies by Zhou~\cite{Zhou2017_1, Zhou2017_2, Zhou2021}.\\
    
    \noindent Recent research has focused on efforts to modify the jet behavior arising from RMIs under shock-loaded conditions~\cite{Sterbentz2022, Jekel2022, Sterbentz2023, Hennessey2023, Kline2024, Schill2024}. Early computational efforts by Sterbentz et al.~\cite{Sterbentz2022} detailed a multi-material buffer designed to modify the growth of RMI in a shock-compressed metal target with sinusoidal features at a metal-air interface. More recently, an experimental effort presented by Kline and Hennessey et al.~\cite{Kline2024} used a single-material, computationally optimized buffer, to suppress the growth of RMIs in an explosively driven linear shaped charge.\\
    
    \noindent In this work, evolution of the interface indicative of RMI was modified in a  copper target containing a conical defect. The copper target was impacted by a detonation wave from an HE driver. Loading was modulated using a multi-material buffer placed between the HE and the target. The work presented here is a computational and experimental extension of the work by Sterbentz et al.~\cite{Sterbentz2022} where an HE driver was employed to impulsively load a metal target, as opposed to a flat shock impactor, and used a copper target with a conical feature to maximize the severity of the RMI. Previous works also focused significant effort on optimizing the design of the buffer, however in this study an experimental design was constructed that demonstrates control over the jet velocity using simple cylindrical buffers composed of two different materials. The buffer dimensions were optimized via simulation using the Lawrence Livermore National Laboratory (LLNL) hydrocode ALE3D~\cite{Noble2017}, and the experiments were carried out and analyzed at the Colorado School of Mines Explosives Research Lab. The validation experiments were performed with a similar set of materials to those used in the simulation work.\\

\end{section}

\begin{section}{Methods}\label{sec:methods}
    \begin{subsection}{Experiment Design}

    To ensure simplicity in design while also demonstrating that jetting in an explosively-loaded copper target can be modulated with a simple buffer, the design for this experiment was chosen to resemble a disc acceleration (DAX) test with some minor modifications~\cite{Lorenz2015}. The experiment involved a cylinder of RDX-based HE (C-4) weakly confined in a plastic tube with an RP-80 detonator, used for initiation in the rear of the tube, and a copper target containing a conical defect in its surface at the front of the tube, shown in Fig.~\ref{fig:configs}(a). In each experiment, a buffer composed of either solid copper, silicone, or a combination of copper and silicone was placed between the HE and the copper target to modify the jetting behavior in the target. The experiment was radially symmetric, allowing for simulation efficiency, as well as ease of experimental data capture via flash X-ray and ultra-high-speed photography. Four buffer configurations were selected: two single-ma,terial buffers served as controls, Fig. \ref{fig:configs}(b)-(c) and two multi-material buffers chosen to either mitigate, Fig.~\ref{fig:configs}(d),  or enhance, Fig.~\ref{fig:configs}(e) the jet velocity in the copper target. The inner radii of the multi-material buffers were hand-optimized such that the modification of the jetting was as significant as possible.
    
\begin{figure}[]
    \centering
    \begin{minipage}[b][][b]{.23\textwidth}
        \centering
        \subfigure[]{\includegraphics[width=0.75\linewidth]{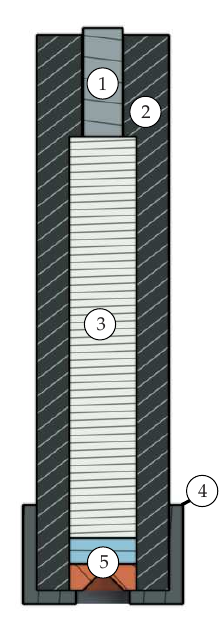}}
    \end{minipage}
    \begin{minipage}[b][][b]{0.38\textwidth}
        \centering\hspace{-2.6em}
      \subfigure[]{\includegraphics[width=0.33\linewidth]{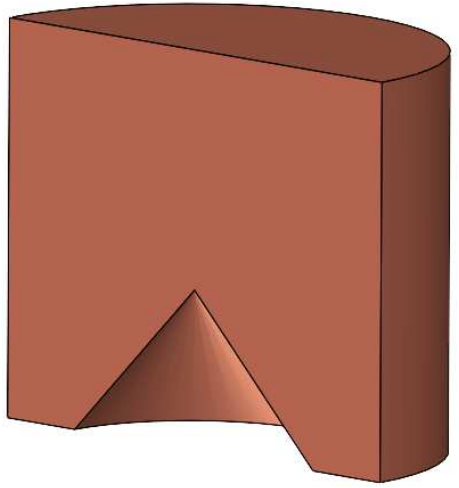}} \hspace{1.65em}
      \subfigure[]{\includegraphics[width=0.33\linewidth]{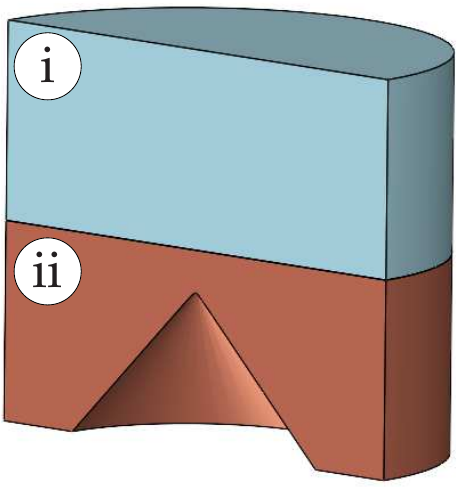}}
      \subfigure[]{\raisebox{4.5mm}{\includegraphics[width=0.33\linewidth]{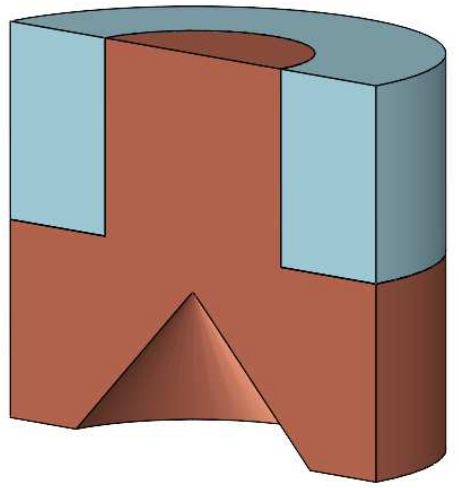}}}
      \subfigure[]{\includegraphics[width=0.62\linewidth]{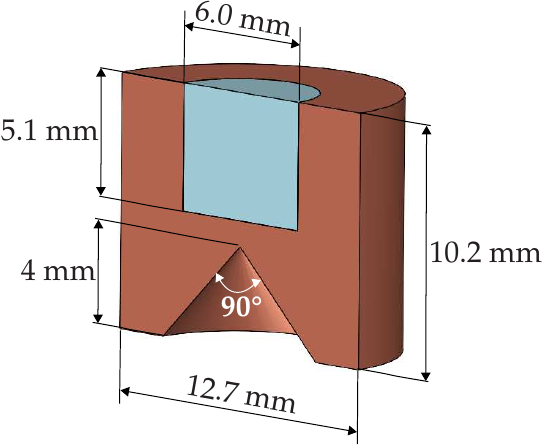}}
    \end{minipage}
        \begin{minipage}[b][][b]{0.30\textwidth}
        \centering \hspace{-2.0em}
      \subfigure[]{\includegraphics[width=0.85\linewidth]{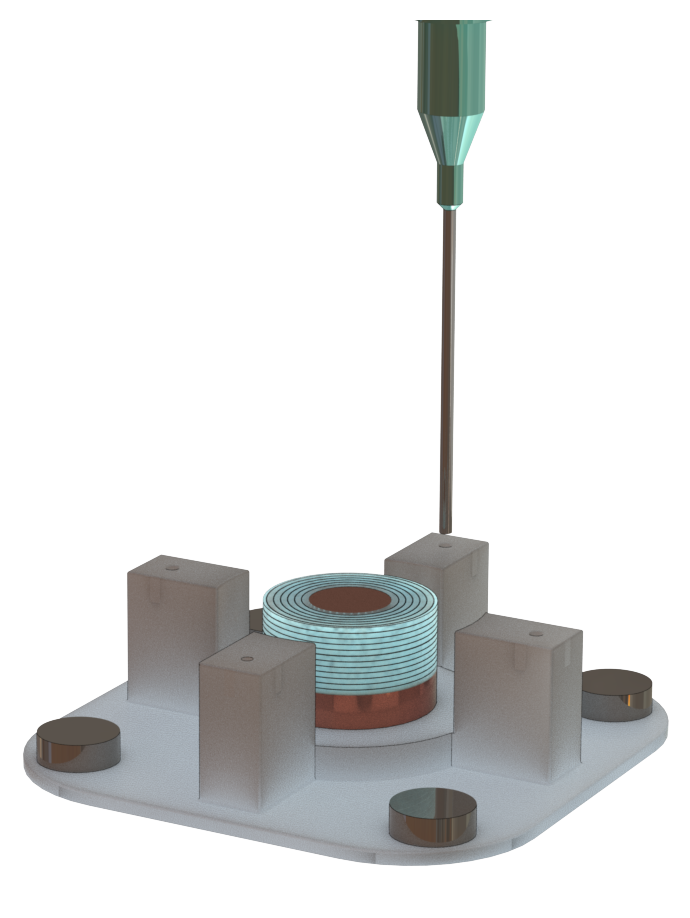}} 
    \end{minipage}
    \label{fig:configs}
\caption{Experimental design of (a) the charge tube with (1) detonator, (2) Delrin charge tube, (3) HE, in this case C-4, (4) 3D printed retaining cap to hold the buffer and target in place, and (5) target and buffer. The four target-buffer configurations that were used in the experimental series: (b) and (c) show the control configurations where (i) denotes copper and (ii) denotes silicone, and (d) and (e) show the mitigating and augmenting jet velocity buffer configurations, respectively. A rendering of the silicone buffer direct ink write 3D printing process is shown in (f).}
\end{figure}

  \end{subsection}

    \begin{subsection}{Materials}

    \noindent Oxygen-free, high-conductivity (OFHC) copper conforming to ASTM B187 grade C10100  ($\leq0.0005$ wt.~\% O) was selected as the target and high-density buffer material. OFHC copper was chosen for its high ductility, abundant equation of state (EOS) information, experimental and computational RMI data, and quality X-ray contrast. Drawn OFHC rods of 12.7~mm-diameter were machined into targets and buffers according to the procedure described in Section 2.3. SE 1700 (Dow), a polydimethylsiloxane elastomer, was used as the low-density silicone buffer material since its dynamic mechanical properties and direct ink write (DIW) 3D printing characteristics are well realized, and highly analogous to Sylgard, which was used as the low-density simulation material. The explosive used in the charges was chosen to be C-4~\cite{MilitaryExplosives1990}. It has an average velocity of detonation of 8046~m/s and a reported Chapman-Jouguet pressure of $1.99\times 10^4$~MPa~\cite{heivilin_production_1990,ahmed2022experimental,kuhl2010thermodynamic}. C-4 was hand-packed into Delrin tubes maintaining a packing density between 1.70 and 1.81~g/cm$^3$. For the simulations, OHFC copper was modeled as copper, Delrin was modeled as Lucite, and SE 1700 was modeled as Sylgard.
    
  \end{subsection}

    \begin{subsection}{Target-Buffer Assembly Fabrication}

    \noindent The copper targets were designed at Lawrence Livermore National Laboratory and manufactured at Colorado School of Mines. OFHC copper rod stock was cut to a thickness of 10.2~mm (0.40~in) and machined on a manual lathe. The conical defect was an inverse cone machined concentrically to a depth of 4.00~mm (0.16~in) at the cone's 90\degree~vertex.  The defect was milled on each target using a 90\degree~end mill mounted onto the tailstock of the lathe. Four copper target-buffers were fabricated. The rear section geometry of each target-buffer varied, where the difference between each configuration was the buffer, a 5.1~mm thick section behind the target.\\
    
    \noindent All configurations of the buffer designs were consistent in thickness and outer diameter, which were 5.1~mm (0.2~in) and 12.7~mm (0.5~in) respectively. The four configurations tested in this study are shown in figure~\ref{fig:configs}. The two control configurations had a buffer section comprising entirely of copper~\ref{fig:configs}(b), or SE 1700~\ref{fig:configs}(c), while the mitigating~\ref{fig:configs}(d) and augmenting~\ref{fig:configs}(e) configurations featured both high- and low-density materials on the back of the copper target. The mitigating configuration, Fig.~\ref{fig:configs} (d) is a 5.1~mm thick section consisting of a 6.0~mm-diameter (0.236 in) machined copper center surrounded by a 3D printed SE 1700 annulus. The augmenting configuration, seen in Fig.~\ref{fig:configs} (e), consists of a 6.0~mm diameter SE 1700 center surrounded by a machined copper annulus. The tolerance of the machined components was 0.127~mm (0.005~in). Before 3D printing the SE 1700 components of each buffer configuration, the copper samples were annealed in a quartz tube furnace at 500 \degree C for 30 minutes in an argon environment and cooled slowly over the course of a day  to ''full soft" to remove any work hardening effects that may have altered the surface properties of the machined copper targets.\\
    
    \noindent Colorado School of Mines utilized an in-house DIW 3D printer to deposit SE 1700 directly on the copper targets. The printer was converted from a commercial-off-the-shelf 3D printer (Prusa i3 Mk3S+) as explained in detail in Sevcik et al.~\cite{sevcik_extrusion_2023}. To print accurately onto the copper targets, jigs were 3D printed from polyethylene glycol (PETG) filament to hold the copper buffers in a set location on the print bed, as shown in Fig.~\ref{fig:configs}(f). Magnets were placed at the corners to keep the jig in place on the magnetic print bed.  The printer's nozzle was manually jogged to the jig's alignment holes. The alignment holes are at a known location on the jig and fit the nozzle tightly. One can determine the exact center of the jig, therefore defining the geometric center of the copper target-buffer in the coordinate system of the printer. The print is started from the center of the jig, as all buffer sections are concentric. The SE 1700 was cured for over 24 hours at 60\degree C before use in a detonation experiment.\\

    \noindent The target-buffers were flush fit to the Delrin charge tube with an outer diameter of 25.4~mm and an inner diameter of 12.7~mm. The main charge was comprised of C-4, which was manually packed into each tube. The tubes held charges of 17.0~$\pm$~0.5~g in intimate contact with both the back side of the buffer and an RP-80 exploding bridgewire (EBW) detonator (Teledyne RISI). The RP-80 was glued into a smaller-diameter hole in the tube, as illustrated in Fig.~\ref{fig:configs}(a). After the remaining cavity had been packed with C-4, the wave shaping buffers were pressed into the cavity and glued to the tube with cyanoacrylate. The bottom end of the tube was fitted with a threaded 3D printed retaining cap to hold the bottom face of the copper flush with the bottom of the tube. Finally, the Delrin tube was inserted into a 3D printed fixture shown in Fig.~\ref{fig:pit} which held the tube in the view of the ultra-high-speed camera and flash X-ray system.\\ 

  \end{subsection}

    \begin{subsection}{Diagnostics}
     
    \noindent The experiments performed at Colorado School of Mines' Explosives Research Laboratory utilized flash X-ray radiography and ultra-high-speed photography. A layout of the experiment in the outdoor explosives arena can be found in Fig.~\ref{fig:pit}. A Shimadzu HPV-X2 ultra-high-speed camera operated in conjunction with a MegaSun illumination system (Prism Sciences) in a point-and-shoot configuration to observe the development of the jet from the target-buffer's conical defect. The MegaSun illumination system consisted of two GN-34 xenon flash lamps. This configuration produced a lighting pulse of approximately 700~$\mu$s duration. The illumination outlasted the detonation event and provided ample exposure for the ultra-high-speed camera. Depending on cloud cover at the outdoor blast facility, neutral density filters (ratings 0.3 -- 0.6) were used on the camera's Nikon AF Micro-Nikkor 200~mm f/4D IF-ED lens. Care was taken to provide a backdrop to increase the contrast of the jet in the images. Matte black paper was used as a backdrop behind the sample, and matte black paint coated the holding fixtures.\\ 
    
    \noindent Shadowgraphs were taken using a 2-channel 450~kV Flash X-ray system and fine-resolution phosphor computed radiography screens (Carestream Industrex Flex GP) which were digitized with a ScanX Discover. Two pulsers (L3Harris) were each configured with a single remote tube head to capture the detonation event. Each pulser unit triggers independently, allowing for two X-ray exposures which are staggered in time. Taking advantage of the axisymmetric targets and charge tubes, the two remote tube heads were oriented orthogonal to one another to minimize cross exposure of the two X-ray channels. The radiography setup was configured to produce images with an object-plane magnification of two at the image-plane. Penumbral blurring was initially minimalized using 2.74~m as a source-to-film distance and a 2~mm spot-size for each pulser. Later experiments used 3~mm spots in an identical configuration which increased penumbral blurring in the shadowgraphs. Uncertainty in the reported timing of each shadowgraph stems from jitter associated with the triggering of a high-energy pulsed power system as well as the standard deviation of the RP-80's function time.\\
    
    \begin{figure}
        \centering
        \includegraphics[width=0.75\textwidth]{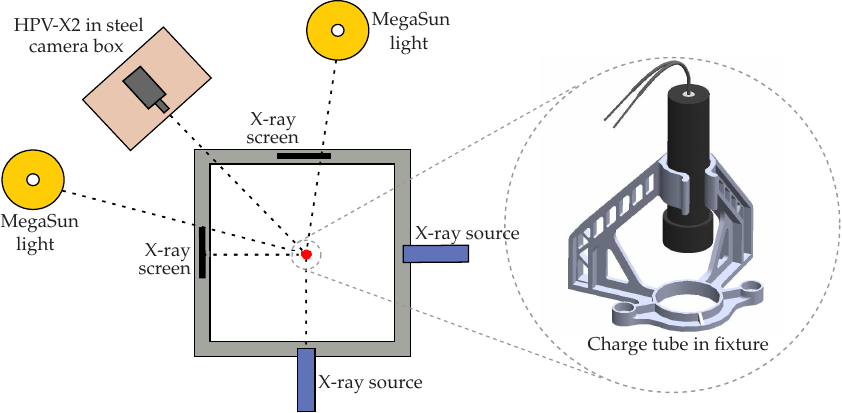}
        \caption{Layout of Colorado School of Mines outdoor explosives arena used for the detonation experiments with flash X-ray and ultra-high-speed diagnostics. A cinder block wall surrounding the charge tube protected all equipment from potential shrapnel. The red dot indicates the location of the charge tube. Not to scale.}
        \label{fig:pit}
    \end{figure}

    \noindent The diagnostic equipment was triggered by a delay generator (Stanford Research Systems model DG645). The lead experimentalist for each shot would trigger the timing array to initiate the experiment. Immediately upon triggering, the MegaSun received a signal which allowed for a $200~\mu$s period to reach peak brightness prior to the detonation event, during which the camera was subsequently triggered. The detonation event was initiated by a capacitor discharge unit (four-channel remote fire-set from Cal-AV labs) triggered by the delay generator. Transit time between the detonator's output face and the interface of the charge with the sample target was estimated using the velocity of detonation and the geometry of the charges. The flash X-ray system was initially set to fire at the time of arrival of the detonation front at the interface. The timing of these shots was delayed by $1~\mu$s in subsequent experimental trials until the resulting shadowgraphs captured dynamic deformation of the targets. Table~\ref{timeline_table} shows start and end times for each piece of equipment. The times are in reference to the experimental time, which starts as the delay generator is triggered. 
    
    \begin{table}[htp]
    \centering
    \caption{Timing for the experimental setup. All times are in reference to the experimental timeline. The time presented for the flash X-ray is a window of time binding the pulsed-event of both exposures}
    \begin{tabular}{|l|c|c|}
    \hline
    \textbf{Equipment} & \textbf{Start time} & \textbf{End time} \\
    \textbf{ } & \textbf{[$\mu${s}]}  & \textbf{[$\mu${s}]} \\ \hline
    \textbf{MegaSun} & 0 & 700 \\ \hline
      \textbf{RP-80} & 200 & 202.65 \\ \hline
     \textbf{HPV-X2} & 202 & 330 \\ \hline
     \textbf{Flash X-ray} & 218 & 224 \\ \hline
    \end{tabular}
    \label{timeline_table}
    \end{table}
    
    \end{subsection}

    \begin{subsection}{Hydrodynamic Simulations and Computational Methods}
    To understand the evolution of the detonation and subsequent jetting behavior, 2D radially symmetric hydrodynamic simulations were performed using the LLNL hydrocode ALE3D, short for arbitrary Lagrangian–Eulerian three-dimensional analysis~\cite{Noble2017}. The simulations were performed at a resolution of 100 zones/cm. The detonator is modeled using a Jones-Wilkins-Lee (JWL) equation of state with a programmed burn detonation model, wherein 5-point detonations are set off in the simulation at equidistant points in the radial direction along the rear of the HE body with a maximum height given by the radius of the detonator. The main HE charge is modeled using inline Cheetah~\cite{Fried1994}. The meaning and units of the parameters used in the following strength model and equation of state are introduced in Table~\ref{tab:params3}.

    \begin{table}
  \centering
    \caption{\label{tab:params3} Description of variables and units used in equations (1), (2) and (3). Spaces left blank denote dimensionless units.}
  \begin{tabular}{|p{0.5in}|p{0.5in}|p{3.0in}|}
    \hline
    Variable & Units & Description \\
    \hline
    $a_n$ & & Binary coefficient describing relative volume dependence on shear modulus\\
    \hline
    $a_p$ & cm$\cdot \mu$s$^2$/g & Pressure dependence of shear modulus\\
    \hline
    $a_T$ & K$^{-1}$ & Linear temperature dependence of shear modulus\\
    \hline
    $b$ & & Linear correction to Gruneisen coefficient\\
    \hline
    $c_0$ & cm/$\mu$s & Sound speed of material at reference condition\\
    \hline
    $E$ & g/cm$\cdot$ $\mu$s$^2$ & Material energy per unit reference volume\\
    \hline
    $G_0$ & g/cm$\cdot$ $\mu$s$^2$ & Reference shear modulus\\
    \hline
    $n$ & & Work hardening exponent\\
    \hline
    $p$ & g/cm$\cdot$ $\mu$s$^2$ & Pressure\\
    \hline
    $S_1$ & & Linear coefficient of shock speed - particle speed relation\\
    \hline
    $S_2$ & & Quadratic coefficient\\
    \hline
    $S_3$ & & Cubic coefficient\\
    \hline
    $T$ & K & Temperature\\
    \hline
    $Y_0$ & g/cm$\cdot$ $\mu$s$^2$ & Reference yield strength \\ 
    \hline
    $Y_\text{max}$ & g/cm$\cdot$ $\mu$s$^2$ & Maximum yield strength \\ 
    \hline
    $\beta$ & & Work hardening coefficient \\ 
    \hline
    $\gamma_0$ & & Reference Gruneisen coefficient \\ 
    \hline
    $\epsilon_0$ & & Strain offset\\
    \hline
    $\epsilon_p$ & & Equivalent plastic strain\\
    \hline
    $\rho$ & g/cm$^3$ & Density \\ 
    \hline
    $\rho_0$ & g/cm$^3$ & Reference density\\
    \hline
  \end{tabular}
\end{table}

    The copper in the target-buffer is modeled using the Steinberg-Guinan strength model given by
    \begin{equation}
      Y = Y_0f(\epsilon_p)\frac{G(p,T)}{G_0},\qquad \text{with} \qquad Y_0f(\epsilon_p) = Y_0\left[1+\beta(\epsilon_p+\epsilon_0)\right]^n\leq Y_\text{max},
    \end{equation}
where
\begin{equation}
  G(p,T) = \left[G_0+a_pG_0p\left(1-a_n+a_n\left(\frac{\rho}{\rho_0}\right)^{-1/2}\right)-G_0a_T(T-300)\right],
\end{equation}
    together with a simple stress-based spall model and a Mie-Gr{\"u}neisen equation of state
    \begin{equation}
      p=\frac{\rho_0c_0^2\mu\left[1+\left(1-\frac{\gamma_0}{2}\right)\mu - \frac{b}{2}\mu^2\right]}{\left[1-(S_1-1)\mu-S_2\frac{\mu^2}{\mu+1}-S_3\frac{\mu^3}{(\mu+1)^2}\right]^2} + (\gamma_0+b\mu)E, \quad \text{with} \quad \mu = \frac{\rho}{\rho_0}-1.
    \end{equation}
    The spall strength parameters are listed in Table~\ref{tab:strParams}. The spall model in use detects whether the maximum principal stress exceeds a value of 1.2 GPa. If spall is detected, the deviatoric stresses are set to zero in that numerical zone. The following Mie-Gr{\"u}neisen equation of state parameters were used for the copper part: $\rho_0=8.93$, $c_0=0.394$, $\gamma_0=2.02$, $b=0.47$, $S_1=1.489$, $S_2=0$, and $S_3=0$. The Lucite case is modeled using a Steinberg-Guinan strength model (Table~\ref{tab:strParams}) and a Mie-Gr{\"u}neisen equation of state with the following parameters $\rho_0=1.182$, $c_0=0.218$, $\gamma_0=0.85$, $b=0$, $S_1=2.088$, $S_2=-1.124$, and $S_3=0$. The SE 1700 portion of the buffer is modeled as Sylgard using a tabular equation of state \cite{Coe2015}.

    \begin{table}[ht]
      \centering
      \caption{Steinberg-Guinan strength model parameters.}
      \begin{tabular}{|l|c|c|c|c|c|c|c|c|c|}
        \hline
        Material & $Y_0$ & $\beta$ & $\epsilon_0$ & $n$ & $Y_\text{max}$ & $G_0$ & $a_p$ & $a_n$ & $a_T$\\ \hline
        Copper & $4.0\times10^{-5}$ & 1200 & 0 & 0.8 & $6.4\times10^{-3}$ & 0.477 & 2.83 & 1 & $3.77\times10^{-4}$\\ \hline
        Lucite & $4.2\times10^{-3}$ & 0 & 0 & 1 & $4.2\times10^{-3}$ & 0.0232 & 0 & 0 & 0\\\hline
      \end{tabular}
      \label{tab:strParams}
\end{table}

    \noindent Tracer particles were placed on the surface of the copper target at the center of the conical defect and at a radius of 0.85~cm out from the center to measure the velocity of the jet. Additionally, a reference velocity was measured along the surface of the copper target, in the opposite direction of the defect. 
    \end{subsection}
\end{section}

\begin{section}{Results and Discussion}\label{sec:results}

    \begin{subsection}{Hydrodynamic Simulations}
    As each article designed for this experiment contains a buffer, the dimensions of the buffer are prime for further exploration, especially when attempting to tailor the desired performance outcome of each buffer configuration. For the single material buffers, which are acting as controls, it is preferable for the performance of each buffer to be as close to identical as possible. For the multi-material buffers, the performance of each buffer was intentionally designed to be as different as possible, such that one buffer maximizes jet velocity and the other minimizes jet velocity. For this purpose, the article was simulated with buffer thicknesses of $t_b= 0.25,0.50,0.75,1.00$~cm, with an arbitrary choice of the internal radius of the layered buffers, taken to be half the radius of the total article. The results of this study showed that the optimal buffer lies between 0.50~cm, and 0.75~cm. Figure~\ref{fig:buffThick} shows plots of jet velocity with different buffer thicknesses. A buffer in this range of thicknesses appears to have enough mass to significantly alter the development of the jet in the copper target in the multi-material case, while allowing the two single-material buffers to act as controls. Due to the realities of machining the buffers and target, choosing a buffer that is identical in thickness to the target simplifies the manufacturing of the parts needed for the experiment significantly. Although this is not the optimal buffer thickness, it is sufficient to show the desired changes in jetting behavior.\\
    \begin{figure}[]
      \centering
      \subfigure[]{\includegraphics[width=0.49\textwidth]{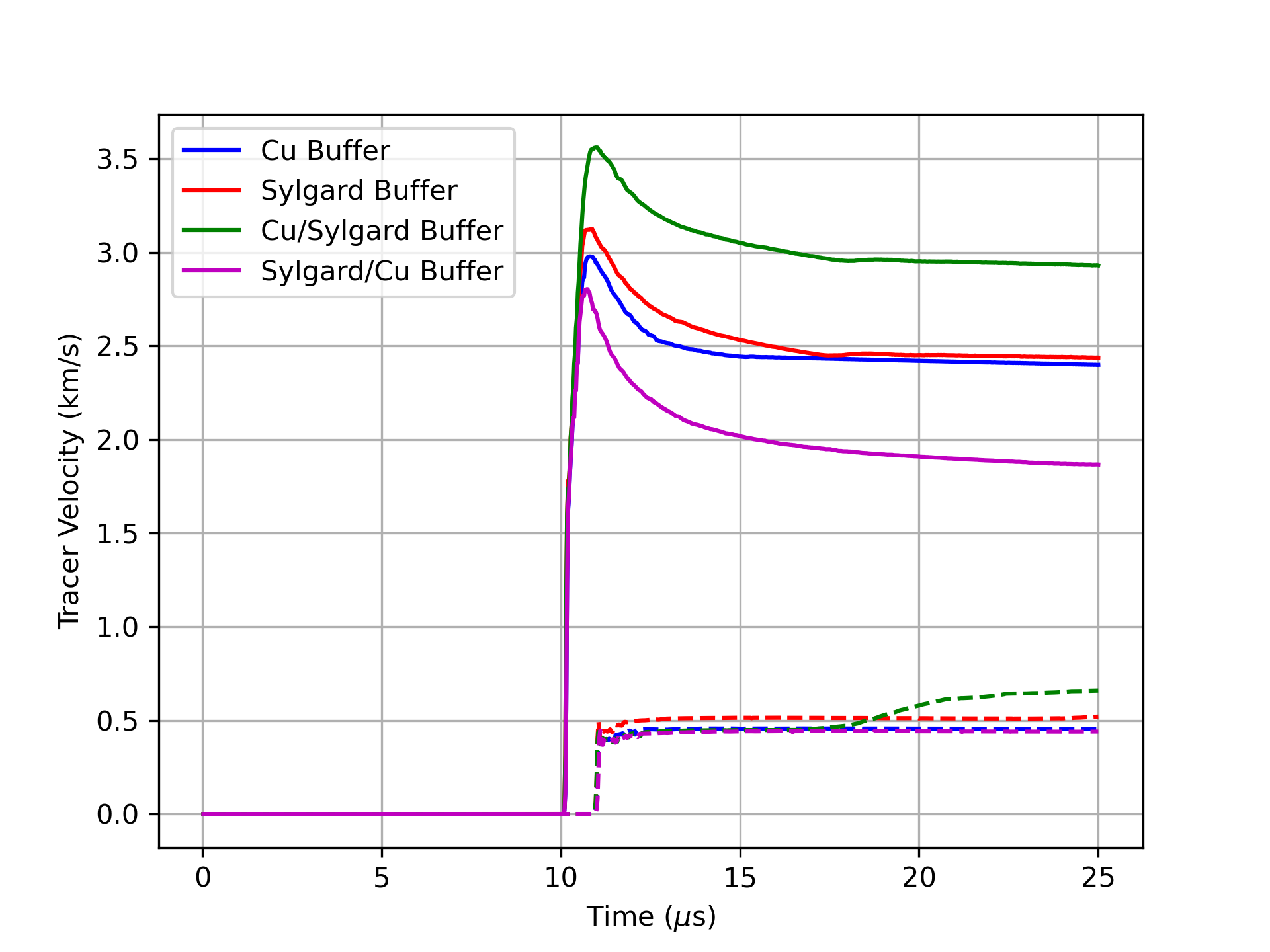}}
      \subfigure[]{\includegraphics[width=0.49\textwidth]{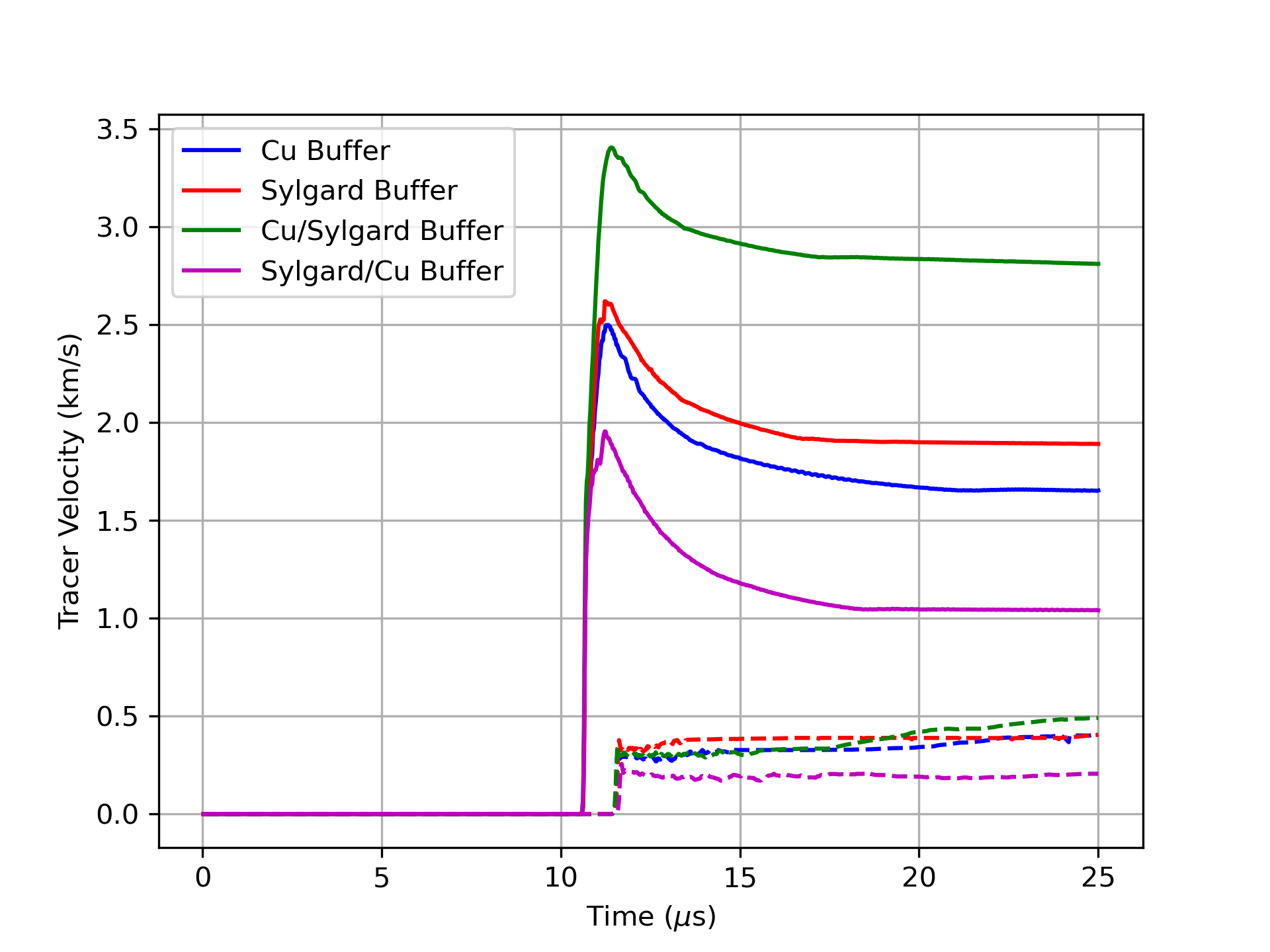}}
      \subfigure[]{\includegraphics[width=0.49\textwidth]{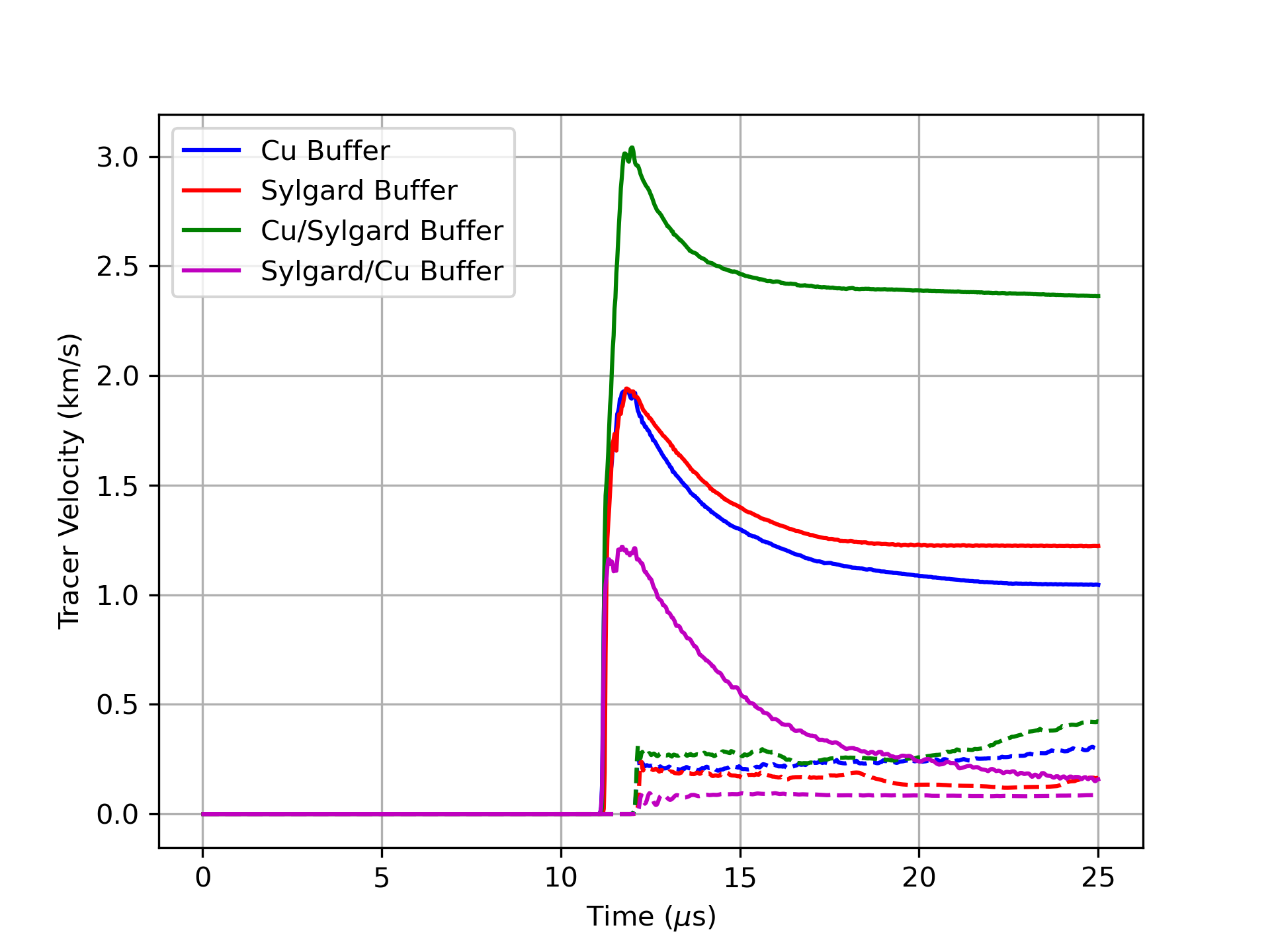}}
      \subfigure[]{\includegraphics[width=0.49\textwidth]{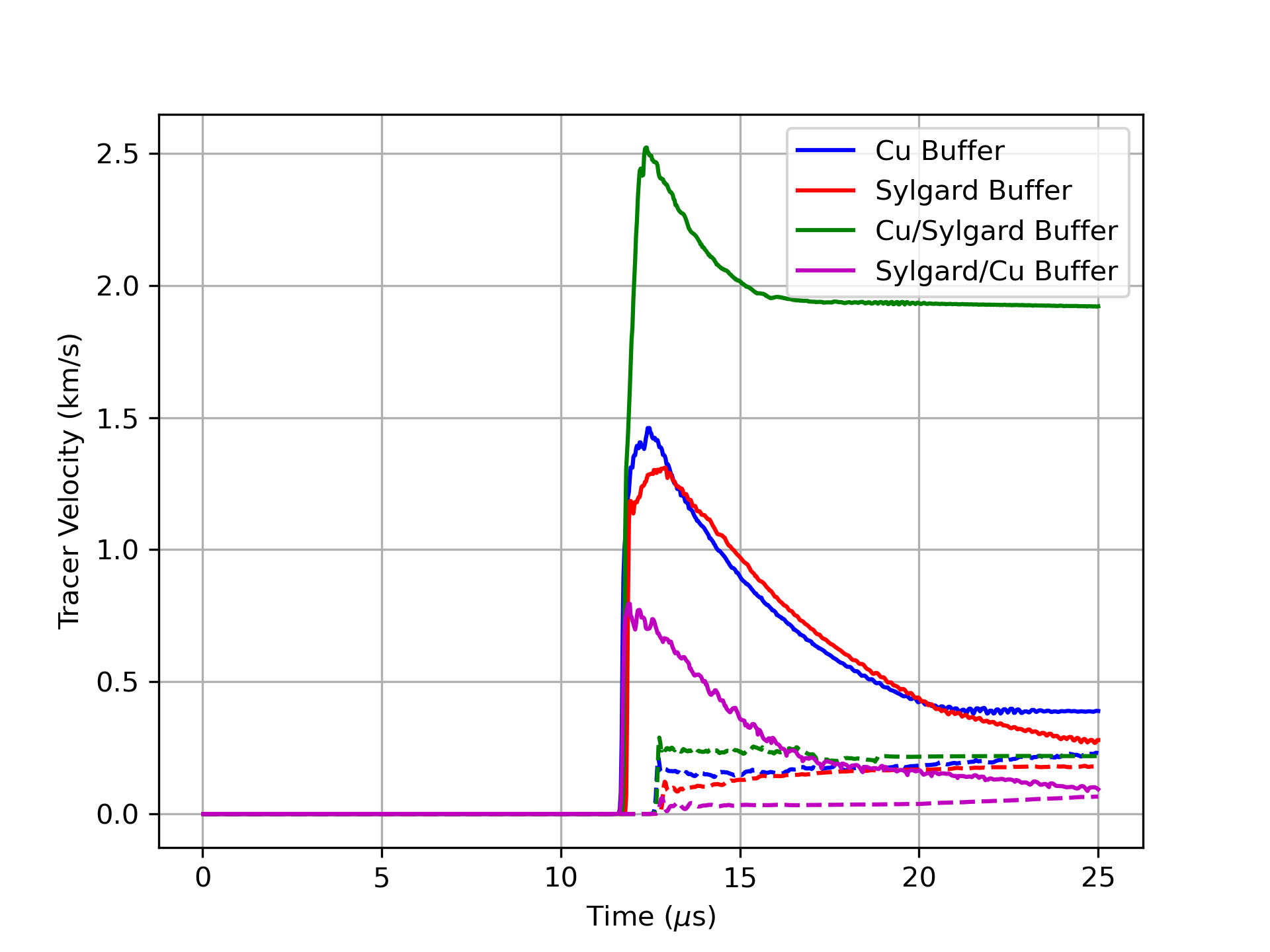}}
      \caption{Tracer velocities for different buffer thicknesses, $t_b$: (a) $t_b=0.25$~cm, (b) $t_b=0.50$~cm, (c) $t_b=0.75$~cm, and (d) $t_b=1.00$~cm.}
      \label{fig:buffThick}
    \end{figure}

    \noindent The second buffer dimension that must be chosen is the internal layer radius for the multi-material buffers. This dimension was chosen in much the same way as the previous. A hand-tuned parameter sweep was performed over the range 0.07~cm to 0.42~cm, with an optimal internal layer radius found to be 0.30~cm. Figure~\ref{fig:buffLayer} shows the jet velocities produced by these final buffer designs. 

    \begin{figure}[!htbp]
      \centering
      \includegraphics[width=0.75\textwidth]{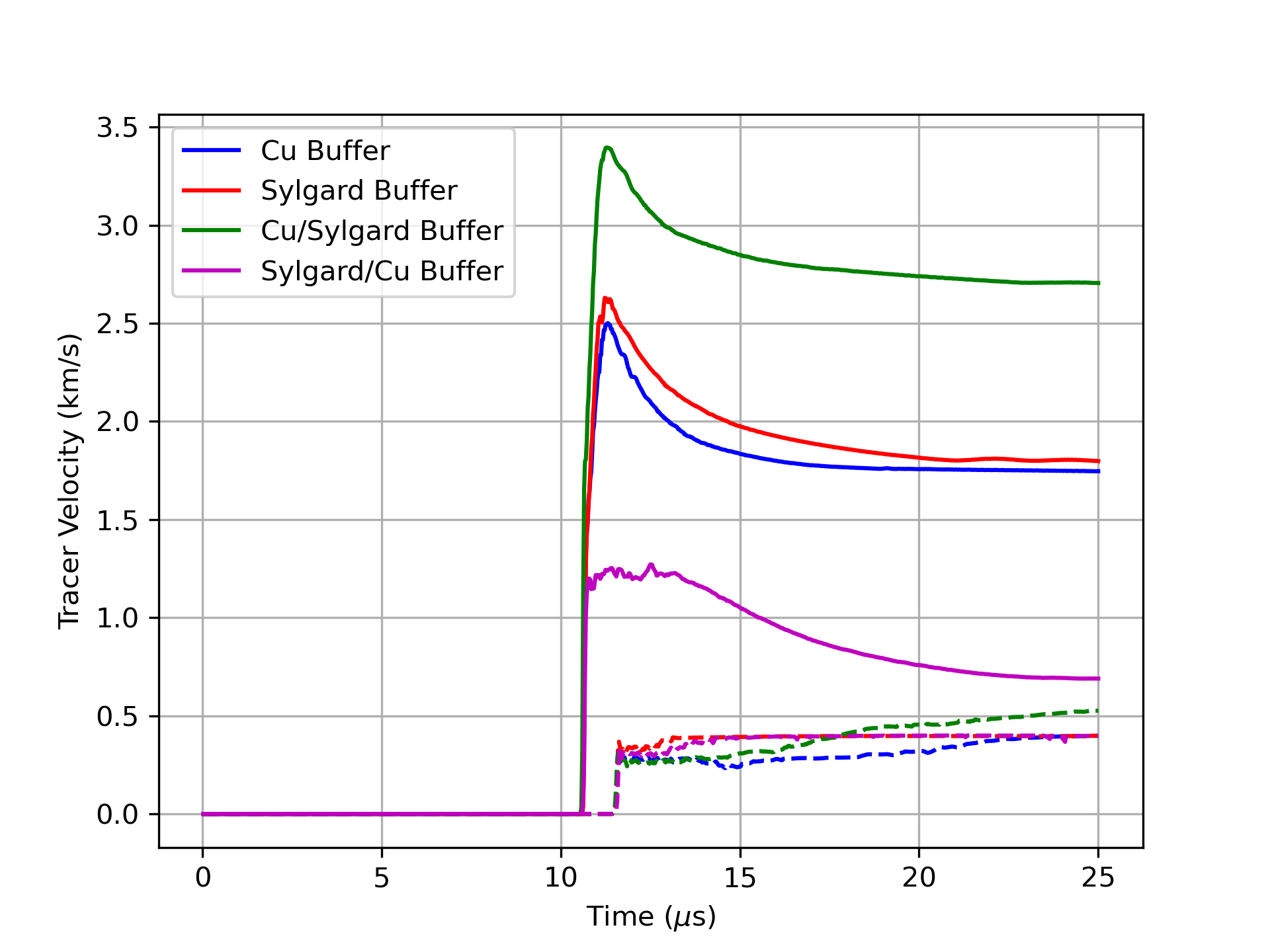}
      \caption{Tracer velocities for the optimal buffer design with thickness 0.50 cm and internal layer radius 0.30~cm.}
      \label{fig:buffLayer}
    \end{figure}

    \noindent As evidenced by the velocity traces in the previous figures, the mechanistic action behind the velocity change in the jet is controlled primarily by the geometry of the buffer, rather than the material used in the buffer. For the single material buffers, the jetting velocity is nearly the same regardless of material used. To explain this, the pressure within the buffer and target was closely examined immediately following the interaction of the detonation wave with the buffer for both article configurations, Fig.~\ref{fig:buff1Evo}. It should first be noted that the shock speed within each single material buffer is nearly identical, 5.16 km/s in Cu vs. 5.06 km/s in Sylgard. 
    The pressures within the buffers themselves are very different, with the shock pressure in the copper buffer reaching nearly 30~GPa while the shock pressure in the silicone buffer is about 10~GPa. However, when the shock refracts at the interface between the buffer and the target, pressures are seen near 30~GPa at that interface when utilizing the silicone buffer. In the pure copper article, the pressure has dissipated somewhat by the time the shock reaches the target to about 26~GPa. Note there is no shock refraction in that case, as the buffer and target are identical materials. Although there is significant difference in pressure, the duration of the pressure pulse on the conical defect in the target must be considered. As can be seen in Fig.~\ref{fig:buff1Evo}, the pressure pulse in the article with the copper buffer is significantly longer than the silicone buffer. It is likely the length of the pressure pulse in the copper buffer case that makes up for the lower pressure peak, ultimately resulting in near identical jetting behavior.\\

    \begin{figure}[!htbp]
      \centering
      \subfigure[]{\includegraphics[height=0.85\textwidth]{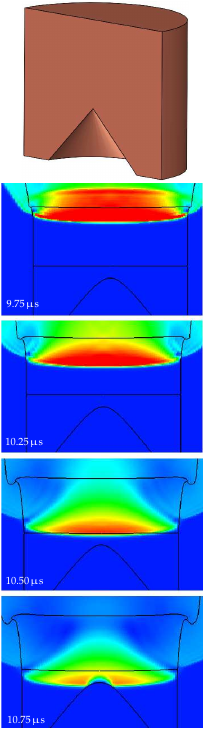}}\hspace{0.1\textwidth}
      \subfigure[]{\includegraphics[height=0.85\textwidth]{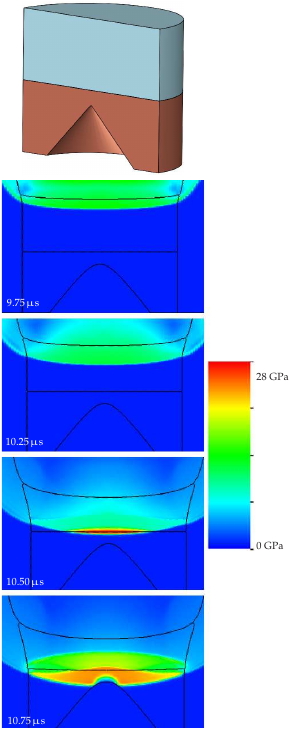}}
      \caption{Time evolution snapshots of the pressure in the article with a (a) copper buffer and (b) silicone buffer. Evolution proceeds from top to bottom at time instants $t=9.75, 10.25, 10.5, \text{~and~} 10.75~\mu$s.}
      \label{fig:buff1Evo}
    \end{figure}

    \noindent The multi-material buffers modify the jetting behavior through geometric effects which combine the mechanistic actions found in the single material buffers. Time evolution snapshots for these buffers can be seen in Fig.~\ref{fig:buff3Evo}. For both of multi-material buffers, the pressure profiles within each buffer section behave much like they do for the single material buffers. Within the silicone portion of the buffer, the pressure drops significantly following the interaction with the detonation wave, while the copper portion of the buffer maintains relatively high pressures. The shock in the silicone portion of the buffer refracts at the copper-silicone interface, resulting in extremely high pressures. The geometric location of the high pressure region following the shock refraction at said interface then influences the development of the jet. When the high pressure region is behind the center of the defect, the jet velocity is increased substantially, whereas if the high pressure region is behind the flat portion of the target the jet velocity is reduced. 

    \begin{figure}[ht]
      \centering
      \subfigure[]{\includegraphics[height=0.85\textwidth]{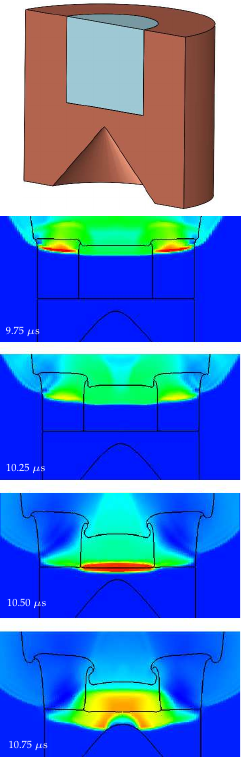}}\hspace{0.1\textwidth}
      \subfigure[]{\includegraphics[height=0.85\textwidth]{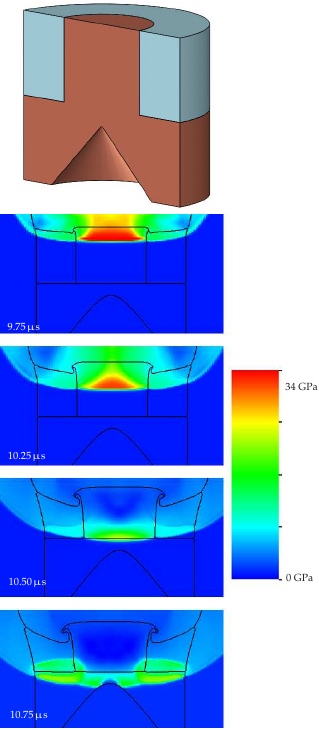}}
      \caption{Time evolution snapshots of the pressure in the article with (a) the jet augmentation buffer and (b) the jet mitigation buffer. Evolution proceeds from top to bottom at time instants $t=9.75, 10.25, 10.5, \text{~and~} 10.75~\mu$s.}
      \label{fig:buff3Evo}
    \end{figure}

  \end{subsection}

    \begin{subsection}{Detonation Experiments}
    The CCD for the HPV-X2 has a size of 400px by 250px. Images are converted into TIFF files and the series of images is imported into ImageJ, allowing the user to set a scale of pixels to millimeters using known dimensions within the frame. The outer diameter of the 3D printed retaining cap at the bottom of the Delrin tube is measured and used as a scale for all experiments. For these experiments, a frame rate of either 2,000,000~fps or 5,000,000~fps was used. A position-time curve was produced by manually tracking the leading edge of the jet from the wave shaping buffer's surface through each frame. The average pixel/mm ratio from the photographs that were processed were $2.74\pm~0.26$. The error of manual image processing of the leading edge is estimated to be $\pm~1.5$~pixels. Figure~\ref{fig:plot2} shows the distance versus time of the various buffer configurations. The data points come from the image processing and a linear fit was applied to the data using the MATLAB R2023a polyfit function. The average slope of the copper control, SE 1700 control, and augmenting buffers were $2.40\pm0.35$~km/s, $2.45\pm0.14$~km/s, and $4.71\pm1.73$~km/s respectively. Perhaps most interestingly, the mitigating buffers did not produce a traceable jet. For the plot, time instance $t= 0$ corresponds to first frame in the image processing series for each configuration and is offset from the experimental or simulation time. Image distortion was not corrected for, as the event being traced is located near the center of the images, where the least amount of distortion is present. Additionally, the lens and filter should introduce minimum distortion at the center of the optical axis.\\

    \noindent The two control configurations, copper and silicone, produced very similar jetting behavior. Figure~\ref{fig:copperImages}~(a)-(c) are several key frames from the HPV-X2 recording. The initial breakout of the bottom of the article is seen at time instance $t=16.39\mu$s, carrying into $t=20.39~\mu$s is assumed to be a jet of gas or powder. By time instance $t=24.39~\mu$s, a solid particulate can be identified just below the bright region at the outlet of the article. The next 20~$\mu$s show a stream of solid particulates. These beads were determined to be the jetting behavior of interest. The plots in Fig.~\ref{fig:plot2} were determined from the leading solid particulate edge in the image series. Deformation of the copper target can be observed between the two shadowgraph images in Fig.~\ref{fig:copperImages}~(d) and Fig.~\ref{fig:copperImages}~(e). The 3-mm spot size of the flash X-ray system created a blurred image without sufficient contrast to observe the jet. The approximate shadowgraph timing can be calculated from the trigger times and function times of the detonator and the pulsers. There is evidence of some reduced grey-scale intensity below the copper target indicating inversion of the defect and possible jetting by time instance $t=23.54\mu$s, which is supported by the high-speed imaging.\\
    
    \noindent The copper control images in Fig.~\ref{fig:copperImages} can be directly compared with the image set of the SE 1700 control buffer in Fig.~\ref{fig:siliconImages}. Due to changes in sunlight, the exposures in this image set are different than the copper control set. The initial breakout from the bottom of the article is shown at time instance $t=20.89\mu$s. Due to the over-exposure of the jet, no individual particulate(s) could be resolved. The leading edge of the illuminated jet body was traced to obtain average velocity. No other particulates were observed to pass the jet tip by time instance $t=38.39~\mu$s. In the case of the pure copper target, the bright gas that is seen at initial break out is passed by the solid jet within $8\mu$s. Figure~\ref{fig:siliconImages}~(e) clearly indicates inversion of the indent and production of a downward facing spike. This spike appears within $1.1\mu$s of the breakout seen in Fig.~\ref{fig:siliconImages}(a), further supporting the evaluation of the jet as the bright structure in the images.\\
    
    \begin{figure}[!htbp]
        \centering
\subfigure[20.39~$\mu$s]{{\includegraphics[width=0.30\textwidth]{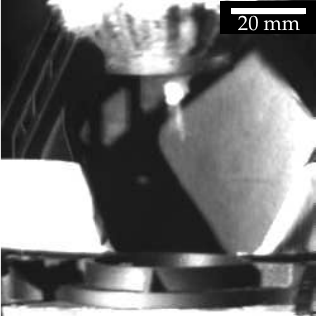}}}
    \subfigure[24.39~$\mu$s]{{\includegraphics[width=0.30\textwidth]{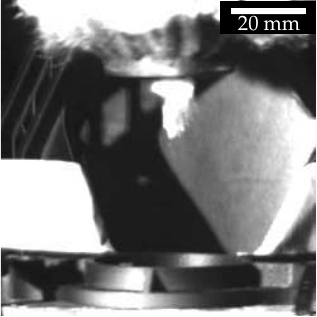}}}
    \subfigure[36.39~$\mu$s]{{\includegraphics[width=0.30\textwidth]{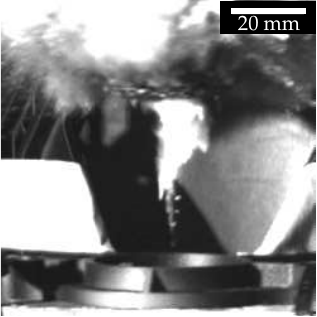}}}
        \subfigure[Copper control target-buffer, static image, pre-shot]{{\includegraphics[width=0.32\textwidth]{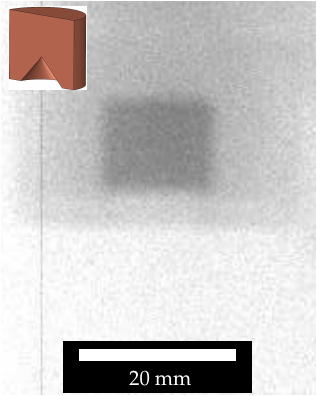}}}
       \subfigure[t$_0$ + 23.54~$\mu{s}$]{{\includegraphics[width=0.32\textwidth]{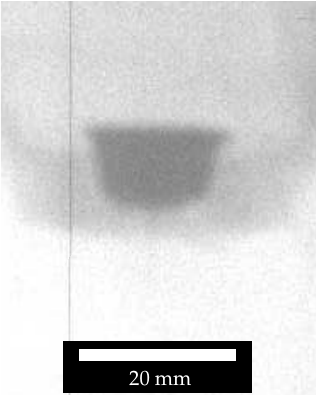}}}
        \caption{Experimental results from the copper control buffer. Photographs recorded at 2 million frames per second are shown in (a)-(c) and shadowgraphs of are shown in (d) and (e).The buffer in configuration is shown in the top left of (d).}%
        \label{fig:copperImages}
    \end{figure}

    \begin{figure}[!htbp]
        \centering
        \subfigure[20.89~$\mu{s}$]{{\includegraphics[width=0.30\textwidth]{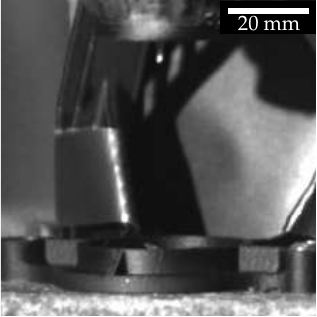}}}\subfigure[30.89~$\mu{s}$]{{\includegraphics[width=0.30\textwidth]{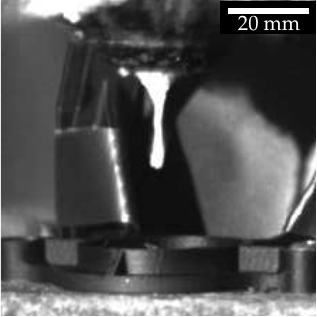}}}\subfigure[38.39~$\mu{s}$]{{\includegraphics[width=0.30\textwidth]{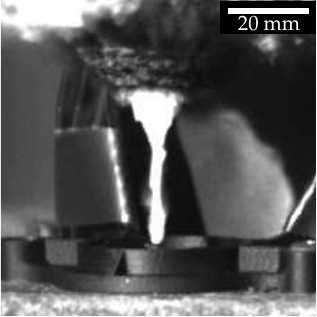}}}
         \subfigure[Silicon control target-buffer, static image, pre-shot]{{\includegraphics[width=0.32\textwidth]{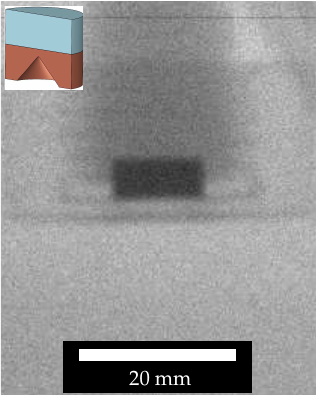}}}
        \subfigure[t$_0$ + 19.54~$\mu{s}$]{{\includegraphics[width=0.32\textwidth]{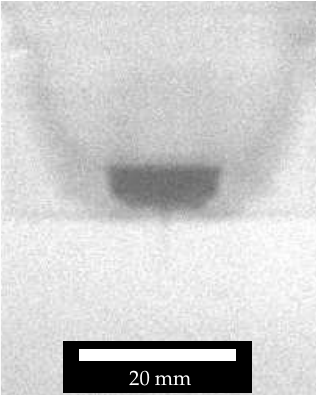}}}
        \caption{{Experimental results from the SE 1700 control buffer. Photographs recorded at 2 million frames per second are shown in (a)-(c) and shadowgraphs of are shown in (d) and (e). The buffer in configuration is shown in the top left of (d).}}%
        \label{fig:siliconImages}
    \end{figure}

    \noindent Figure~\ref{fig:mitigatingImages}~(a)-(c) shows key frames selected from the ultra-high-speed photographs during a shot with the mitigating buffer configuration. Destruction of the 3d printed retaining cap can be observed from $t=19.89~\mu$s to $t=29.89~\mu$s, indicating that the detonation front has arrived and passed the target-buffer. No jet or other ejecta phenomenon are observed from this experiment. Fig.~\ref{fig:mitigatingImages}~(d) and Fig.~\ref{fig:mitigatingImages}~(e) shows two dynamic shadowgraphs taken during the detonation experiment. No visible inversion of the target occurs. Similar deformation is seen in the simulation of the mitigating buffer in Fig.~\ref{fig:buff3Evo}, approximately $10~\mu$s earlier than the shadowgraphs, indicating that the detonation wave has since passed and most deformation has already occurred.\\ 
    
    \noindent The results from the augmenting buffer configuration are shown in Fig.~\ref{fig:augmentingImages}. The initial breakout seen at $t=19.89~\mu$s develops into an off-axis jet of gas or powder. This artifact dissipates by $t=27.39\mu$s. The solid particulate jet can be seen emerging at $t=24.89\mu$s, and can be traced from there. There is evidence that the jet had already began to form as indicated in the shadowgraphs in Fig.~\ref{fig:augmentingImages}~(e) and Fig.~\ref{fig:augmentingImages}~(e) and Fig.~\ref{fig:augmentingImages}~(f). By $t=19.54~\mu$s, a spike structure has formed after the target defect has inverted.  The plot in Fig.~\ref{fig:plot2} shows two augmenting cases with considerable run to run variability. This is likely a consequence of using manually-packed explosives charges. Regardless of the spread between the two runs, both results shown an increased jet velocity compared to the control cases.\\                                                
    
    \begin{figure}
        \centering\subfigure[19.89~$\mu{s}$]{{\includegraphics[width=0.30\textwidth]{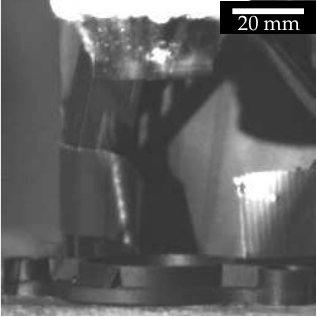}}}\subfigure[24.89~$\mu{s}$]{{\includegraphics[width=0.30\textwidth]{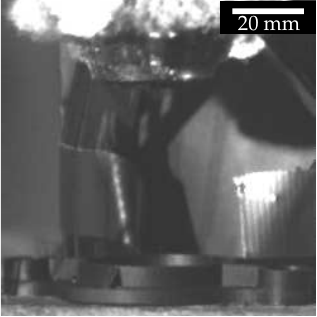}}}\subfigure[29.89~$\mu{s}$]{{\includegraphics[width=0.30\textwidth]{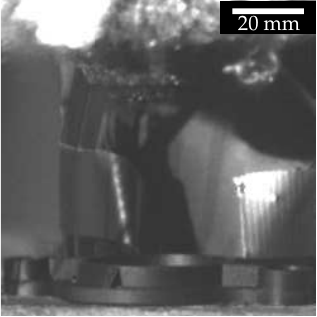}}}
        \subfigure[Mitigating target-buffer, static image, pre-shot]  {{\includegraphics[width=0.32\textwidth]{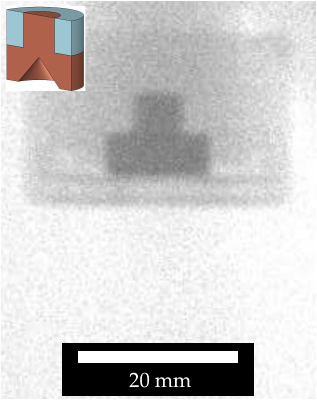}}}
        \subfigure[t$_0$ + 19.54~$\mu{s}$]{{\includegraphics[width=0.32\textwidth]{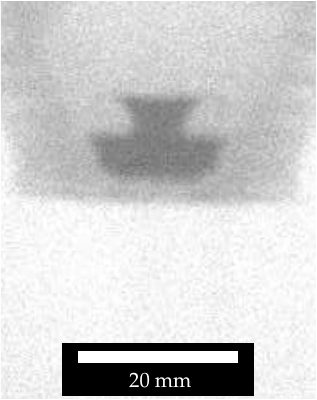}}}
        \subfigure[t$_0$ + 22.54~$\mu{s}$]{{\includegraphics[width=0.32\textwidth]{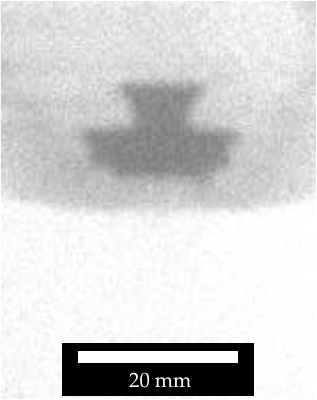}}}
        \caption{Experimental results from the mitigating buffer. Photographs recorded at 2 million frames per second are shown in (a)-(c) and shadowgraphs of are shown in (d) through (f).The buffer in configuration is shown in the top left of (d).}
    \label{fig:mitigatingImages}
    \end{figure}

    \begin{figure}
        \centering\subfigure[19.89~$\mu{s}$]{{\includegraphics[width=0.30\textwidth]{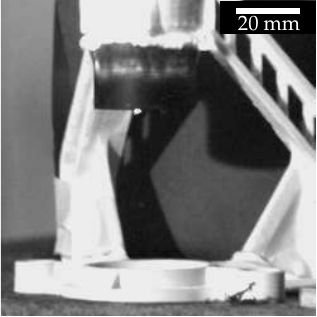}}}\subfigure[24.89~$\mu{s}$]{{\includegraphics[width=0.30\textwidth]{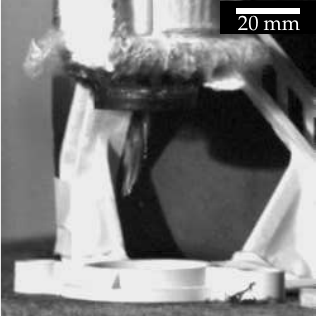}}}\subfigure[27.39~$\mu{s}$]{{\includegraphics[width=0.30\textwidth]{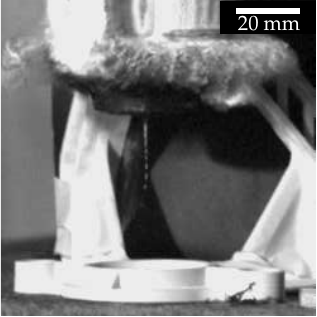}}}
         \subfigure[Augmenting target-buffer, static pre-image]{{\includegraphics[width=0.32\textwidth]{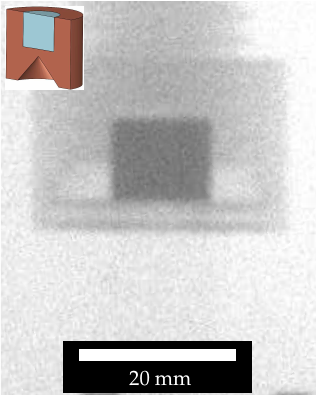}}}
        \subfigure[t$_0$ + 19.54~$\mu{s}$]{{\includegraphics[width=0.32\textwidth]{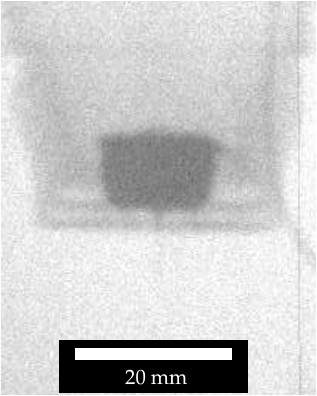}}} 
        \subfigure[t$_0$ + 22.54~$\mu{s}$]{{\includegraphics[width=0.32\textwidth]{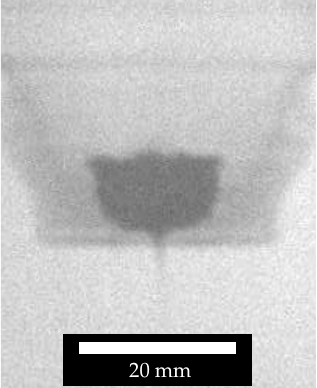}}}
        \caption{Experimental results from the augmenting buffer.Photographs recorded at 2 million frames per second are shown in (a)-(c) and shadowgraphs of are shown in (d) through (f).The buffer in configuration is shown in the top left of (d).}
    \label{fig:augmentingImages}
    \end{figure}

\begin{figure}[!htbp]
        \centering\includegraphics[width=0.75\textwidth]{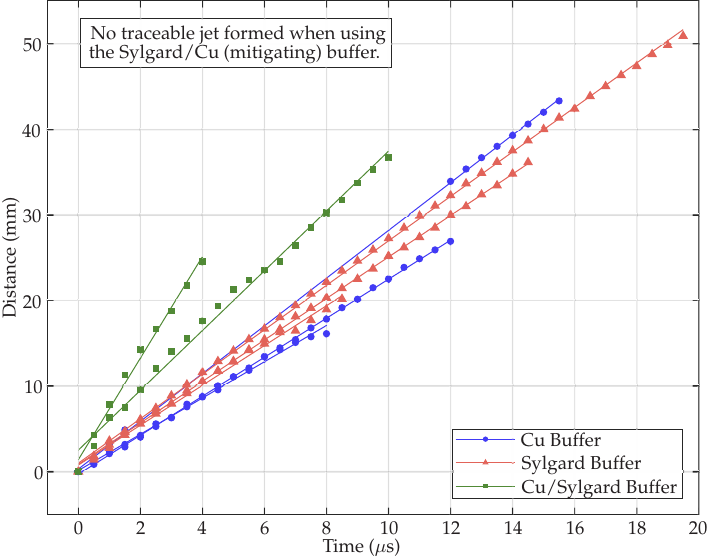}
        \caption{Distance versus time results for the jet tip of the various buffer configurations. Note that the jet velocity mitigating case data is not presented as there was no visible jet in the captured flash X-ray or ultra-high-speed images.} 
        \label{fig:plot2}
    \end{figure}
    
    \end{subsection}
\end{section}

\begin{section}{Conclusions}\label{sec:conclusions}
    \noindent In this work, the control over jet velocity in an explosively loaded copper target with a conical defect using simple cylindrical multi-material buffers has been demonstrated. The mechanistic action responsible for the effects of the multi-material buffers was discussed, and detonation experiments show jet velocity modification in agreement with the simulation results. The mechanistic action discussed in this work is considerably different than the mechanism behind RMI reduction in the previously mentioned study \cite{Sterbentz2022}. This implies that the mechanism behind jet velocity modification is a function of the driven impulse within the metal target. Future work may be designed around using these mechanistic actions to modify jetting in alternative configurations.
\end{section}

\clearpage
\begin{section}{CRediT Author Statement}

\textbf{Michael P. Hennessey}: Conceptualization, Methodology, Software Validation, Formal Analysis, Investigation, Resources, Data Curation, Writing - Original Draft, Writing - Review \& Editing, Visualization, Supervision. 
\textbf{Finnegan Wilson}: Conceptualization, Methodology, Validation, Formal Analysis, Investigation, Resources, Data Curation, Writing - Original Draft, Writing - Review \& Editing, Visualization. 
\textbf{Grace I. Rabinowitz}: Conceptualization, Methodology, Validation, Formal Analysis, Investigation, Resources, Data Curation, Writing - Original Draft, Writing - Review \& Editing, Visualization. 
\textbf{Max J. Sevcik}: Conceptualization, Methodology, Validation, Investigation, Resources Writing - Original Draft, Writing - Review \& Editing, Visualization. 
\textbf{Kadyn J. Tucker}: Methodology, Validation, Investigation, Resources Writing - Original Draft, Writing - Review \& Editing. 
\textbf{Dylan J. Kline}: Conceptualization, Methodology, Validation, Formal Analysis, Investigation, Resources, Data Curation, Writing - Original Draft, Writing - Review \& Editing, Visualization, Supervision, Project Administration. 
\textbf{David K. Amondson}: Conceptualization, Methodology, Validation, Visualization, Supervision, Project Administration. 
\textbf{H. Keo Springer}: Conceptualization, Software, Validation, Formal Analysis, Investigation, Resources, Visualization, Supervision, Project Administration. 
\textbf{Kyle T. Sullivan}: Conceptualization, Methodology, Software, Validation, Investigation, Writing - Review \& Editing, Visualization, Supervision, Project Administration, Funding Acquisition. 
\textbf{Veronica Eliasson}: Conceptualization, Methodology, Validation, Formal Analysis, Investigation, Data Curation, Writing - Original Draft, Writing - Review \& Editing, Visualization, Supervision, Project Administration, Funding Acquisition. 
\textbf{Jonathan L. Belof}: Conceptualization, Methodology, Validation, Investigation, Writing - Review \& Editing, Visualization, Supervision, Project Administration, Funding Acquisition. 

\end{section}

\begin{section}{Acknowledgements}

\noindent Authors Michael P. Hennessey and Finnegan Wilson contributed equally to this work.\\

\noindent This work was performed under the auspices of the U.S. Department of Energy by Lawrence Livermore National Laboratory under Contract DE-AC52-07NA27344. We gratefully acknowledge the LLNL Lab Directed Research and Development Program for funding support of this research under Project No. 21-SI-006. Document release number LLNL-JRNL-860841-DRAFT. 
\end{section}

\clearpage
\bibliography{1-article-LLNLmines}
\end{document}